\documentclass[runningheads]{svmult}

\usepackage{makeidx}   % allows index generation
\usepackage{graphicx}  % standard LaTeX graphics tool
\usepackage{subeqnar}  % subnumbers individual equations
\usepackage{multicol}  % used for the two-column index
\usepackage{physprbb}  % modified textarea for proceedings,
\makeindex             % used for the subject index
\begin{document}
\title*{%
Spectroscopy of PNe in Sextans A, Sextans B,  NGC 3109 and Fornax
}
\toctitle{Spectroscopy of PNe in Sextans A, Sextans B,  NGC 3109 and Fornax}
% allows explicit linebreak for the table of content
%
%
\titlerunning{Spectroscopy of PNe in Sextans A, Sextans B,  NGC 3109 and Fornax}
% allows abbreviation of title, if the full title is too long
% to fit in the running head
%
\author{Alexei Y.\ Kniazev\inst{1}
\and Eva K.\ Grebel\inst{2}
\and Alexander G.\ Pramskij\inst{3}
\and Simon A.\ Pustilnik\inst{3}}
\authorrunning{Alexei Kniazev et al.}
% if there are more than two authors,
% please abbreviate author list for running head
%
%
\institute{Max-Planck-Institut f\"ur Astronomie,
K\"onigstuhl 17, D-69117 Heidelberg, Germany
\and
Astronomisches Institut, Universit\"{a}t Basel, Venusstrasse 7,
CH-4102 Binningen, Switzerland
\and
Special Astrophysical Observatory, Nizhnij Arkhyz,
Karachai-Circassia, 369167, Russia
}

\maketitle              % typesets the title of the contribution

\begin{abstract}
Planetary nebulae (PNe) and HII regions provide a probe of the chemical
enrichment and star formation history of a galaxy from intermediate ages
to the present day. Furthermore, observations of HII regions and PNe
permit us to measure abundances at different locations, testing the
homogeneity with which heavy elements are/were distributed within a galaxy.
We present the first results of NTT spectroscopy of HII regions and/or PNe
in four nearby dwarf galaxies: Sextans A, Sextans B, NGC 3109, and Fornax.
The first three form a small group of galaxies just beyond the Local
Group and are gas-rich dwarf irregular galaxies, whereas Fornax is a
gas-deficient Local Group dwarf spheroidal that stopped its star
formation activity a few hundred million years ago. For all PNe
and some of the HII regions in these galaxies we
have obtained elemental abundances via the classic $T_{\rm e}$-method based
on the detection of the [OIII] $\lambda$4363 line.
The oxygen abundances in three HII regions of Sextans A
are all consistent within the individual rms uncertainties.
The oxygen abundance in the PN of Sextans A is however significantly higher.
This PN is even more enriched in nitrogen and helium, implying
its classification as a PN of Type I.
The presumably unaffected PN abundances of S and Ar are well below those in the
HII regions, indicating a lower metallicity at the epoch of the PN
progenitor formation. For two HII regions in Sextans B, the oxygen
abundances do not differ within the rms uncertainties.
The third one is, however, twice as metal-rich, providing
evidence for the inhomogeneity of the current metallicity distribution
in Sextans B. For the PN in Sextans B we measured an O/H that is
consistent with that of the low-metallicity HII regions.
For NGC~3109 our preliminary results indicate that the oxygen abundances
of PNe and HII regions are all within a small range of $\pm$0.15 dex.
For the PN in Fornax, Ne, Ar and S abundances suggest
that the ISM metallicity was $\sim$0.3 dex lower at the epoch of
the PN progenitor's formation, compared to the O/H value derived for the PN.
\end{abstract}

\section{Introduction}

Understanding how the elemental abundances of galaxies have changed over
time is an essential issue of galaxy evolution studies.
Abundance measurements constrain theoretical models, providing
important clues to how galaxies evolve. In particular, combining the data
on PNe numbers and on PN and HII region elemental abundances can allow one to
derive an approximate enrichment and star formation
history of a galaxy from intermediate ages
to the present day. While HII regions indicate the present-day gas-phase
elemental abundances ($\sim10$ Myr), PNe, although they cannot be accurately
age-dated,  reveal the chemical composition of a galaxy at ``intermediate''
ages of a few 100 Myr to a few Gyr.
Furthermore, if HII regions and PNe at different locations are studied,
this permits us to test the
homogeneity with which heavy elements are/were distributed within a galaxy.
For the Local Group and other nearby galaxies these data can be combined with
star formation histories derived from color-magnitude
diagrams of resolved stars, thereby yielding deeper insights on galaxy evolution.
The goal of our present work was to improve our understanding of metallicities
both at the current epoch (from HII regions), and in
previous periods of star formation (from PNe) on the
basis of new high-quality NTT spectrophotometry for three members of
Antlia-Sextans group of dwarf galaxies  \cite{vdB99}
and for the Fornax dwarf spheroidal galaxy.

\section{Observations and Reduction}

Spectrophotometric observations of HII regions and PNe in the targeted galaxies
were conducted with the NTT at ESO, La Silla, in February 2004.
The observations were performed with the Red Arm of
the EMMI multipurpose instrument with a long slit of
8$^{\prime} \times$ 2$^{\prime\prime}$.
For these observations the CCD rows were binned by a factor of 2,
yielding a final spatial sampling of 0.33$^{\prime\prime}$ pixel$^{-1}$.
The seeing during the observations was very stable and varied from night
to night in the range of 0.4$^{\prime\prime}$ to 0.6$^{\prime\prime}$.
The whole spectral range covered by the two grisms was of 3800 -- 8700 \AA\
with the sampling of 1.6 \AA\ pixel$^{-1}$ for the blue (3800 -- 7000 \AA)
and 1.4 \AA\ pixel$^{-1}$ for the red part (5750 -- 8670 \AA) of the spectra.
Each galaxy was observed with slit positions which covered several
HII regions and also one PN in Sextans A, one PN in Sextans B,
four PNe in NGC 3109 and one PN in Fornax.
The PNe positions have been taken from \cite{Magr03} for Sextans A,
from \cite{Magr02} for Sextans B,
from \cite{RM92} for NGC~3109 and from \cite{Danz78} for Fornax

All primary reductions were done using the IRAF and MIDAS environments.
The method of emission line measurements and the
calculation of element abundances was described in detail in
\cite{Izotov94,IT99,Kniazev00,Kniazev03,Kniazev04}.
For the majority of PNe and HII regions in these galaxies O, N, S, Ar and He
abundances were obtained with the classic $T_{\rm e}$-method after
the detection of the [OIII] $\lambda$4363 line.

\begin{figure}[t]
\begin{center}
\includegraphics[angle=-90,width=0.80\textwidth]{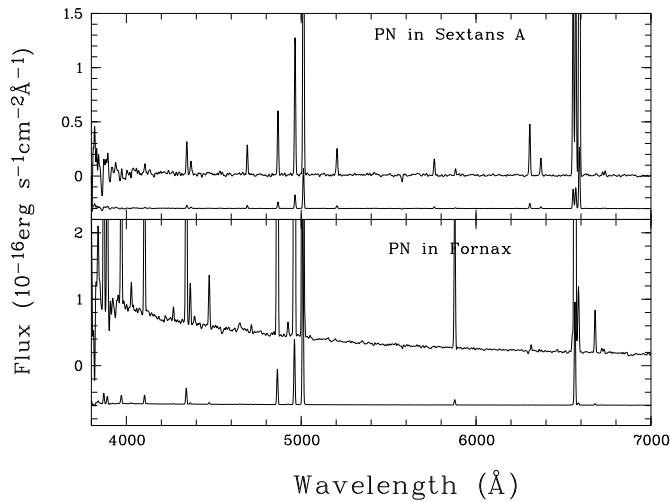}
\end{center}
\caption[]{%
Planetary nebulae in Sextans A and Fornax.  The emission-line EMMI/NTT
spectra were obtained with grism \#5 and cover a wavelength range of
3800 and 7000 \AA.
The spectra at the bottom of each panel is scaled by 1/10 (top) and 1/30
(bottom) and shifted to show the relative intensities of strong lines.
}
\label{eps1}
\end{figure}

\section{PN abundances vs. HII abundances}

HII region metallicities mainly provide information about abundances
of $\alpha$-process elements, produced predominantly in short-lived massive stars.
In contrast to HII regions, some elemental abundances
in PNe are affected by nucleosynthesis in PN progenitors.
It is well known that newly synthesized material can be dredged up
by convection
in the envelope, significantly altering abundances of He, C and N in the
surface layers during the evolution of 1--8 M$_\odot$ stars on the giant branch
and asymptotic giant branch (AGB) \cite{LD96,BS99,HKB00}.
In addition, if during the thermally pulsing phase of AGB evolution, convection
``overshoots'' into the core, significant amounts of $^{16}$O can
be mixed into
the inter-shell region and may be convected to the surface \cite{Her00,Bl01}.
In combination, these factors mean that surely only Ne, S and Ar abundances,
observed in both HII regions and PNe, can be used to probe the enrichment history
of galaxies.

\section{Sextans A}

Sextans A is a dIrr at a distance of 1.32 Mpc for which deep HST
images have permitted the measurements of stars to a limiting absolute magnitude of
$M_{\rm V} \sim$ +1.9 \cite{Dolphin03}. The gas abundance of heavy elements
for Sextans A was first determined by \cite{SKH89}.
This study presented only an empirical
estimate of O/H, which has a characteristic rms uncertainty of $\sim$50\%.
While this estimate was successfully used in various statistical studies
to combine the data on O/H with the CMD-derived SF history and metallicities,
such accuracy is not sufficient.

We have derived an average value of 12+log(O/H)=7.57 from three
HII regions in Sextans A.
The values of N/O, S/O and Ar/O are consistent in all three
HII regions to within their rms observational uncertainties and are close to those
derived for the subgroup of the most metal-poor HII galaxies \cite{IT99}.
New, high accuracy chemical abundances for HII regions in
Sextans A allow us to probe metallicity homogeneity across
the body of the galaxy. For the three HII regions observed in Sextans A,
the measured abundances show no differences exceeding 0.1 dex.
Moreover, these metallicities are in good agreement with those in
three A-supergiants in Sextans A \cite{Kaufer04}.

Chemical abundances derived for the PN in Sextans A
show that this is a Type I object (see Figure~1),
with a highly elevated nitrogen abundance: N/O $\sim$2.
Its 12+log(O/H) = 8.02 is about a factor of 3 higher than in the HII regions,
which implies significant self-pollution by the PN progenitor.
Since the abundances of S and Ar should not be altered in a PN progenitor
(see discussion above),
their values indicate that the ISM metallicity was $\sim$0.5 dex lower
at the epoch of the PN progenitor's formation,
compared to the current metallicity in HII regions and A-supergiants.

\section{Sextans B}

Sextans B is a dIrr at a distance of 1.36 Mpc \cite{Kar02}, in which
red giants, intermediate-age stars, and young stars are found.
As far as its global properties are concerned, Sextans B
is considered to be a ``twin'' to Sextans A, but Sextans B is of particular
interest because of a discrepancy in oxygen abundances between
the HII measurements of \cite{SCV86,SKH89,MAM90}.

We confirmed with good accuracy the low ISM metallicity of Sextans B
with 12+$\log$(O/H) = 7.53$\pm$0.04 in two HII regions.
The element abundance ratios of O, N, S and Ar are well
consistent with the respective patterns of very metal-poor
HII galaxies. We have found that one HII region is significantly enriched, with
an excess of O, N, S and Ar abundances, relative to the mean value
of two other HII regions, by a factor of 2.5$\pm$0.5.
The elemental abundances of the observed PN in Sextans B are consistent with
those of the two HII regions with the low metallicity.

\section{NGC~3109}

NGC 3109, at a distance of 1.33 Mpc, is the most massive dIrr galaxy in
Antlia-Sextans group.
Compared to galaxies of similar total luminosity (e.g., the SMC),
the optical extent is approximately two times larger.
The H I gas, stellar content and star formation history of this galaxy
have been studied, e.g., by \cite{Car85,GMTF93}.
The underlying old stars are metal-poor ([Fe/H] = $-$1.7 dex), as
measured by \cite{Grebel03}.
\cite{Lee03} reported an oxygen abundance for one HII region with
the $T_{\rm e}$-method, although the S/N in the [OIII] $\lambda$4363 line
was low ($\sim$~2.5).
Our preliminary measurement of O/H for this HII region is quite consistent
with their value,
but with higher S/N=6 for the [OIII] $\lambda$4363 line.
We found that oxygen abundances of PNe and HII regions are all in a narrow
range near 12+log(O/H) $\sim$ 7.65$\pm$0.15.

\section{PN in Fornax}

The PN in the Fornax dwarf spheroidal galaxy was the only one found
by \cite{Danz78}, who were the first to observe it spectroscopically.
Our data are considerably better (see Figure~\ref{eps1}), yielding
S/N=25 for the [OIII] $\lambda$4363 line.
We have calculated 12+log(O/H) = 8.28 for this PN,
which is slightly lower than the O/H value 8.38 from \cite{Maran84}
that was calculated on the basis of line intensities from \cite{Danz78}.
With our new data we have for first time detected weak
[SII] $\lambda\lambda$6717,6731 lines (I(6717+6731) $<$ 0.015 I(H$\beta$))
and have determined an electron number density, N$_e$(SII) = 750 cm$^{-3}$.
With the newly determined O, N, Ne, Ar and S abundances
and knowledge of mean heavy-element abundance ratios for these elements
from \cite{IT99} we have found that the PN is enriched with oxygen and
the ISM metallicity
was $\sim$0.3 dex lower at the epoch of the PN progenitor formation
compared to the value O/H derived for the PN. This result is
consistent with the chemical evolution scenario for the Fornax dwarf
derived by \cite{Tol03}.
The Fornax PN shows a Wolf-Rayet blue bump in the spectrum that provides
important constraints on the central star evolutionary status.
It is worth noting that both PNe from the Sagittarius dwarf spheroidal galaxy
from \cite{Walsh97} also show Wolf-Rayet features in their spectra.

%INDEX
% Please check with the editor of your book whether he plans to
% include a "mutual" subject index - if so, please code your entries
% in the standard syntax. For your own purposes you may print your
% "personal" index by using the following commands:
%
%\clearpage
%\addcontentsline{toc}{section}{Index}
%\flushbottom
%\printindex
%%


\begin{thebibliography}{35.}
\addcontentsline{toc}{section}{References}

\bibitem{Bl01} Bl\"ocker, T. 2003, in IAU Symp. 209, Planetary Nebulae:
	       their evolution and role in the Universe,
	       eds. S.Kwok, M.Dopita, \& R.Sutherland, 101

\bibitem{BS99} Boothroyd, A.I., \& Sackmann, I.-J. 1999, ApJ, 510, 232

\bibitem{Car85} Carignan, C. 1985, ApJ, 299, 59

\bibitem{Danz78} Danziger, I.J., Dopita, M.A., Hawarden, T.G., \& Webster, B.L. 1978, ApJ, 220, 458

\bibitem{Dolphin03} Dolphin, A., Saha, A., Skillman, E.  et al. 2003, AJ, 126, 187

\bibitem{Grebel03} Grebel, Gallagher, \& Harbeck 2003, AJ, 125, 1926

\bibitem{GMTF93} Greggio, L., Marconi, G., Tosi, M., \& Focardi, P. 1993, AJ, 105, 894

\bibitem{Her00} Herwig, F. 2000, A\&A, 360, 952

\bibitem{HKB00} Henry, R.B.C., Kwitter, K.B., \& Bates, J.A. 2000, ApJ, 531, 928

\bibitem{Izotov94} Izotov, Y.I., Thuan, T.X., \& Lipovetsky, V.A. 1994, ApJ, 435, 647

\bibitem{IT99} Izotov, Y.I., \& Thuan, T.X. 1999, ApJ, 511, 639

\bibitem{Kar02} Karachentsev, I.D., Sharina, M.E., Makarov, D.I., et al. 2002, A\&A, 389, 812

\bibitem{Kaufer04} Kaufer, A., Venn, K.A., Tolstoy, E., Pinte, C., \& Kudritzki, R.-P.
	2004, AJ, 127, 2723

\bibitem{Kniazev00} Kniazev, A.Y., Pustilnik, S.A., Masegosa, J., et al. 2000,
A\&A, 357, 101

\bibitem{Kniazev03} Kniazev, A.Y., Grebel, E.K., Hao, L., Strauss, M.A.,
Brinkmann, J., \& Masataka Fukugita 2003, ApJ, 593, L73

\bibitem{Kniazev04} Kniazev, A.Y., Pustilnik, S.A., Grebel, E.K.,
Lee, H., \& Pramskij, A.G. 2004, ApJS August 2004, v153 issue, in press, (astro-ph/0404133)

\bibitem{Lee03} Lee, H., McCall, M.L., Kingsburgh, R.L., Ross, R.,
		\& Stevenson, C.C. 2003, AJ, 125, 146

\bibitem{LD96} Leisy, P., \& Dennefeld, M. 1996, A\&AS, 116, 95

\bibitem{Maran84} Maran, S.P., Gull, T.R., Stecher, T.P., Aller, L.H., \&
		  Keyes C.D. 1984, 280, 615

\bibitem{Magr02} Magrini, L., Corradi, R.L.M., Walton, N.A., et al. 2002, A\&A, 386, 869

\bibitem{Magr03} Magrini, L., Corradi, R.L.M., Greimel, R., et al. 2003, A\&A, 407, 51

\bibitem{MAM90} Moles, M., Aparicio, A., \& Masegosa, J. 1990, A\&A, 228, 310

\bibitem{RM92} Richer, M.G., \& McCall, M.L. 1992, AJ, 103, 54

%\bibitem{Sakai97} Sakai, S., Madore, B.F., \& Freedman, W.L. 1997, AJ, 480, 589

\bibitem{SKH89} Skillman E.D., Kennicutt, R.C., \& Hodge, P.W. 1989, ApJ, 347, 875

\bibitem{SCV86} Stasinska, G., Comte, G., \& Vigroux, L. 1986, A\&A, 154, 352

\bibitem{Tol03} Tolstoy, E., Venn, K.A., Shetrone, M., et al. 2003, AJ, 125, 707

\bibitem{vdB99} van den Bergh 1999, AJ, 517, L97

\bibitem{Walsh97} Walsh, J.R., Dudziak, G., Minniti, D.
		  \& Zijlstra, A.A. 1997, ApJ, 487,651

\end{thebibliography}
\end{document}